\documentclass[doublecol]{epl2}

\title{Ratio of shear viscosity  to entropy density in multifragmentation of Au + Au}
\shorttitle{Ratio of shear viscosity  to entropy density in multifragmentation of Au + Au}

\author{ C. L. Zhou\inst{1,2}
\and Y. G. Ma \inst{1} \thanks{E-mail: \email{Corresponding author. ygma@sinap.ac.cn}}
\and D. Q. Fang \inst{1}
\and S. X. Li \inst{1,2}
\and G. Q. Zhang \inst{1}
}

\shortauthor{C. L. Zhou \etal}

\institute{
  \inst{1} Shanghai Institute of Applied Physics, Chinese Academy of Sciences,
Shanghai 201800, China\\
\inst{2} Graduate School of
 the Chinese Academy of Sciences, Beijing 100080, China}

 {Viscosity of liquids; diffusive momentum transport}

\abstract{The ratio of the shear viscosity ($\eta$) to entropy density ($s$) for the intermediate energy
heavy-ion collisions has been calculated by using the Green-Kubo method in the framework
of the quantum molecular dynamics model. The theoretical curve of $\eta/s$ as a function of the
incident energy for the head-on Au+Au collisions displays that a minimum region of $\eta/s$ has been
approached at higher incident energies, where the minimum $\eta/s$ value is about 7 times Kovtun-Son-
Starinets (KSS) bound (1/4$\pi$). We argue that the onset of minimum $\eta/s$ region at higher incident
energies corresponds to the nuclear liquid gas phase transition in nuclear multifragmentation.}

\begin{document}

\maketitle

\section{\label{sec:level1}I. INTRODUCTION}

Intermediate energy heavy ion collisions have been extensively studied experimentally
and theoretically for obtaining information about the properties of
nuclear matter under a wide range of density and
temperature. One of the most important aspects of studying
nucleus-nucleus collisions at these conditions focuses on multifragmentation and   liquid gas
phase transition (LGPT) around a hundred  MeV/nucleon
 \cite{Gupta,Bona,Bor,Poc,Nato,Ma-NST,Gil,Ell,Ma-PRL,Gross1990,Bondorf1995}. On the other hand, the ratio of shear viscosity to entropy density ($\eta/s$)  has been claimed to reach its local minimum at the phase transition temperature for a wide class of systems. Empirical observation of the temperature
or incident energy dependence of the shear viscosity to entropy
density ratio  for H$_2$O, He and Ne$_2$ exhibits a minimum in the
vicinity of the critical point for phase transition \cite{Csernai2006}. And a
lower bound of $\eta/s > 1/4\pi$, obtained by Kovtun-Son-Starinets (KSS)
for infinitely coupled super-symmetric Yang-Mills gauge theory based
on the AdS/CFT duality conjecture, is speculated to be universally valid \cite{Kovtun2005,Policastro2001}. In ultra-relativistic heavy-ion collisions \cite{Demir2009,Lacey2007,Chen,Kapu2008,Maj2007,XuZhe2009}, the ratio of shear viscosity to entropy density was used  for studying the quark-gluon plasma phase and  a minimum value of $\eta/s$ close to the lower bounder was claimed. In contrast, the study on $\eta/s$ was very limited in intermediate energy heavy-ion collision. Therefore it is of very interesting to investigate shear viscosity or $\eta/s$ in the intermediate energy domain \cite{Chen2007,Danielewicz1984,Shi2003,Pal2010,Auerbach2009}. In our recent paper~\cite{Li2011}, the relation between $\eta/s$ and incident energy or temperature is studied in the framework of one-body mean-filed theory, namely Boltzmann-Uehling-Uhlenbeck (BUU) model, a gradual decreasing behavior of $\eta/s$ was observed in lower energy but tends saturated around a hundred MeV/nucleon. However no minimum was found, which may be due to the absence of the dynamical fluctuation and cluster formation in the BUU model. In contrast, some other signals of multifragmentation indicate a turning point around  a certain beam  energy or temperature in experimental data as well as in some models, such as  the quantum molecular dynamics (QMD) model. In this context, it will  be very interesting  to investigate the $\eta/s$ in the QMD model.

In this work we use a microscopic transport model known as the isospin-dependent quantum dynamic (IQMD) model \cite{Ma2001}, to simulate Au+Au central collisions. The thermodynamic and transport properties are extracted from the nuclear matter located in the central sphere, with radius $R$=3.5 fm. The generalized hot Thomas Fermi  formula (GHTFF) \cite{khoa1992a,khoa1992b,puri1992} are employed to extract the thermodynamic properties, eg. temperature and  entropy density. The shear viscosity is calculated by using the Green-Kubo relation. Furthermore the multiplicity of intermediate mass fragments (IMFs) is also studied as a signal of liquid gas phase transition \cite{Ma1995} to verify the Green-Kubo's result.

The paper is organized as follows. Section 2 provides a brief introduction of QMD model. In Section 3 we present calculation and discussion. The conclusion and outlook is given in Section 4.

\section{\label{sec:level2} II. QMD model}

The quantum molecular dynamics  model approach is a many-body theory that describes heavy-ion collisions from intermediate to relativistic energy \cite{Aichelin1991} . The isospin-dependent quantum molecular dynamics model \cite{Hartnack1989,Hartnack1998}  is based on the QMD model, including the isospin degrees and Pauli blocking etc.
Each nucleon in the colliding system is described as a Gaussian wave packet

\begin{equation}
\label{eq:gauspack}
  \phi_i({\bf r},t) = \frac{1}{{(2\pi L)}^{3/4}}
exp[-\frac{{({\bf r}- {\bf r_i}(t))}^2}{4L}] exp[-\frac{i{\bf r}
\cdot {\bf p_i}(t)}{\hbar}].
\end{equation}

Here ${\bf r}_{i}(t)$ and ${\bf p}_{i}(t)$ are the mean position and mean momentum, respectively, and the Gaussian width has the fixed value $L$=2.16 fm$^{2}$ for Au + Au system.
The centers of these Gaussian wave packets propagate in coordinate (${\bf R}$) and momentum
(${\bf P}$) space according to the classical equations of motion:
\begin{equation}
\dot{{\bf p}}_{i}=- \frac{\partial \langle {\bf H}
\rangle}{\partial {\bf r}_i}; \dot{{\bf r}}_i=\frac{\partial
\langle {\bf H} \rangle}{\partial {\bf p}_i},
\end{equation}
where $\langle {\bf H} \rangle$ is the Hamiltonian of the system.

The Wigner distribution function for a single nucleon density in phase space is given by

\begin{equation} \label{wignerfuction}
 f_i ({\bf r},{\bf p},t) = \frac{1}{\pi^3 \hbar^3 }
 {\rm e}^{-({\bf r} - {\bf r}_{i} (t) )^2  \frac{1}{2L} }
 {\rm e}^{-({\bf p} - {\bf p}_{i} (t) )^2  \frac{2L}{\hbar^2}  }.
\end{equation}

The mean field in IQMD model is:
\begin{equation}
U(\rho) = U_{\rm Sky} + U_{\rm Coul}  + U_{\rm Yuk} + U_{\rm sym},
\end{equation}
 where $U_{\rm Sky}$, $U_{\rm Coul}$, $U_{\rm Yuk}$, and $U_{\rm
sym}$  represents the Skyrme potential, the Coulomb potential, the
Yukawa potential and the symmetry potential interaction,
respectively \cite{Aichelin1991}.
The Skyrme potential is:
\begin{equation}
U_{\rm Sky} = \alpha(\rho/\rho_{0}) + \beta{(\rho/\rho_{0})}^{\gamma},
\end{equation}
 where $\rho_{0}$ = 0.16
{fm}$^{-3}$ and $\rho$ is the nuclear density. In the present work, the parameter set with
$\alpha$ =-356 MeV, $\beta$ = 303 MeV, and $\gamma$ = 7/6, is used, which
corresponds to a soft equation of state. $U^{\rm Yuk}$ is
a long-range interaction (surface) potential, and takes the
following form

\begin{eqnarray}
 \label{v_yuk}
&U^{Yuk} & =  ({V_y}/{2}) \sum_{i \neq j}{exp(Lm^2)}/{r_{ij}} \nonumber \\
&\cdot & [exp(mr_{ij})erfc(\sqrt{L}m-{r_{ij}}/{\sqrt{4L}}) \nonumber \\
&-& exp(mr_{ij})erfc(\sqrt{L}m+{r_{ij}}/{\sqrt{4L}})]
\end{eqnarray}
 with $V_y =
0.0074$GeV and  $m = 1.25{fm}^{-1}$.
$r_{ij}$ is the relative distance between two nucleons. The
symmetry potential is $U_{sym} = 32 \frac{\rho_n -\rho_p}{\rho_0}\tau_z$,
where $\rho_n$, $\rho_p$, and $\rho_0$ are
the  neutron,  proton and nucleon densities, respectively; $\tau_z$ equals
1 or -1 for neutrons and protons, respectively.

From Eq.~(\ref{wignerfuction}) one obtains the matter density of coordinate space by the sum over all the nucleons, i.e.

\begin{eqnarray}
\label{eqn_rho}
 \rho({\bf r},t)= \sum_{i=1}^{A_{T}+A_{P}}\rho_{i}({\bf r},t)=\sum_{i=1}^{A_{T}+A_{P}} \frac{1}{(2{\pi}L)^{3/2}} e^{-\frac{({\bf r}-{\bf r}_i(t))^2}{2L}}.
\end{eqnarray}

\begin{figure}
\includegraphics[width=8.5cm]{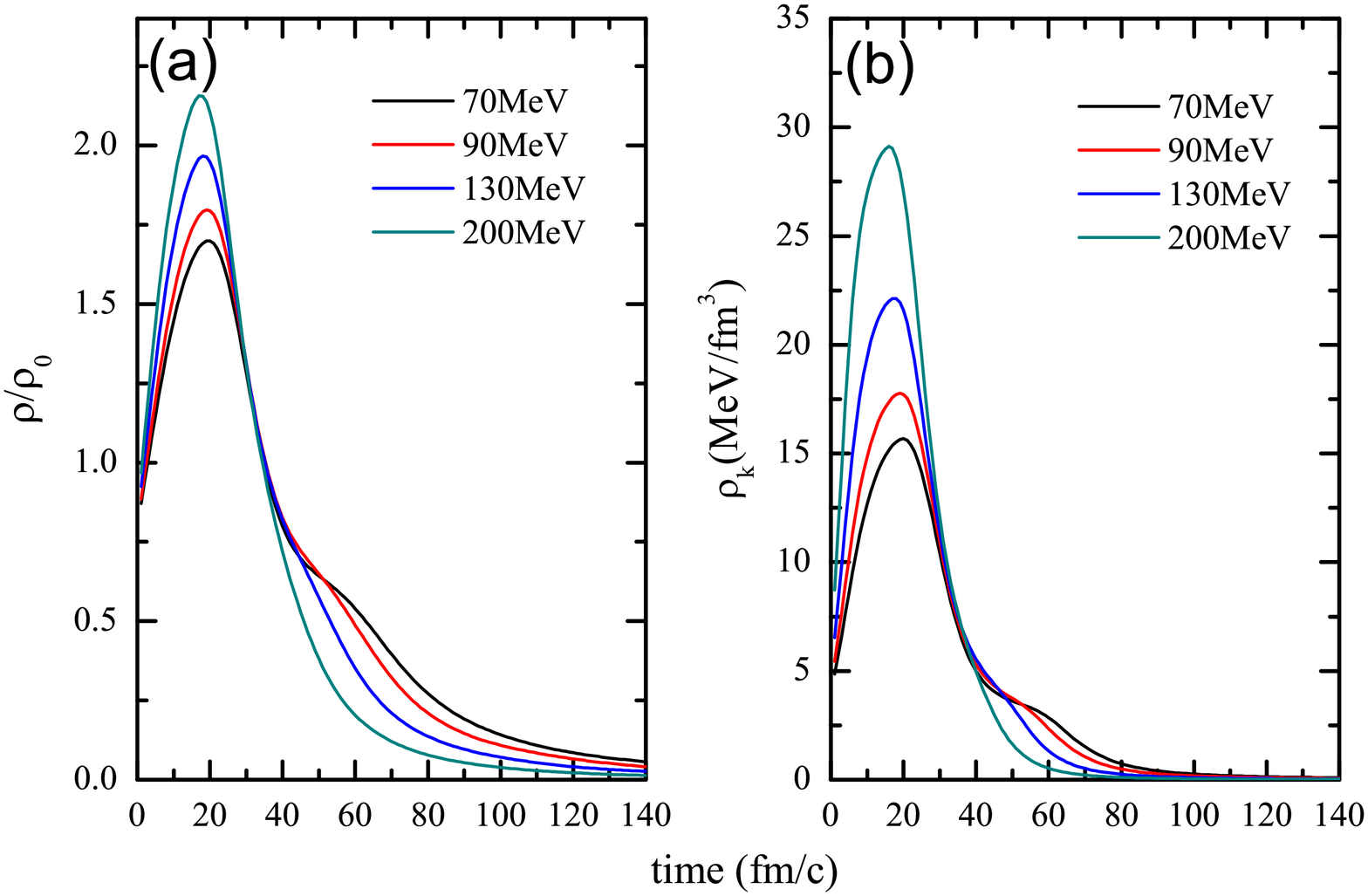}
\vspace{-0.1truein} \caption{\footnotesize The time evolution of mean nuclear matter density  (a) and kinetic energy density (b) in  a central region defined as a sphere  with radius R=3.5fm for the head-on Au+Au collisions. Different color line represents different beam energy which is illustrated in the inset.
}
\label{fig_rhotao}
\end{figure}

The kinetic energy density in coordinate space could also be calculated from Eq. (\ref{eqn_rho})
\begin{eqnarray}
\rho_{K}({\bf r},t)= \sum_{i=1}^{A_{T}+A_{P}}\frac{{{\bf P}_i(t)}^2}{2m}\rho_{i}({\bf r},t).
\end{eqnarray}
The time evolution of the mean nuclear density and kinetic energy density in a given central volume (with $R$ = 3.5 fm) is shown in Fig.~\ref{fig_rhotao}. Both the matter density and kinetic energy density are reaching their maxima around 20 fm/c. And the hot and dense nuclear matter survives for a longer time when the beam energy is lower. This can be easily understood as the nuclear matter experiences compressed and expanded more quickly at higher beam energy.

\begin{figure}
\includegraphics[width=8.5cm]{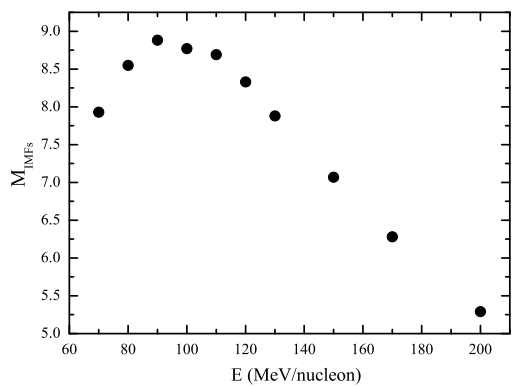}
\vspace{-0.1truein} \caption{\footnotesize   $M_{IMFs}$ as a function of beam energy for the head-on Au+Au  collisions.}
\label{fig_imf}
\end{figure}

With the help of coalescence mechanism, the fragment information can be given in IQMD. The intermediate mass fragment which is here defined  with charge number greater than 3 and smaller than 1/3 of the system size is very important for nuclear multifragmentation. These fragments are larger than typical evaporated light particles and smaller than the residues and fission products, and they can be considered as nuclear fog. So the multiplicity of intermediate mass fragments ($M_{IMFs}$) is  related to the occurrence of liquid gas phase transition. Usually the $M_{IMFs}$  increases first as beam energy increases when the nuclear liquid phase is still dominant, and reaches a maximum, then decreases when the nuclear gas phase becomes dominant \cite{Ma1995}.
Fig. ~\ref{fig_imf} shows the $M_{IMFs}$ which is extracted from the final stage of the collision as a function of beam energy for head-on Au+Au  collisions. One can see the turning energy is around  90 MeV/nucleon. If we study the thermodynamic evolution of $\eta/s$ as a function of
beam energy, it is probably to exhibit a minimum at a certain beam energy, which means $\eta/s$ could also serve as a probe of nuclear liquid gas phase transition in  heavy ion collisions.

\section{\label{sec:level3} III. Calculations and Discussions}
\begin{figure}
\includegraphics[width=8.5cm]{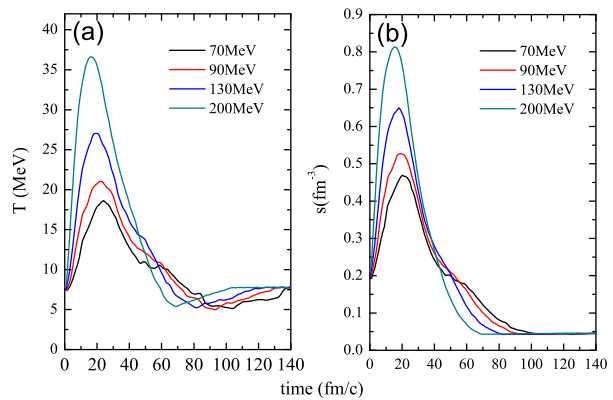}
\vspace{-0.1truein} \caption{\footnotesize   Time evolution of temperature (a) and entropy density (b) in the central region. Different color line represents different beam energy as  illustrated in the inset.
}\label{fig_temend}
\end{figure}

Thermodynamical properties of hot nuclear matter created in heavy ion
collisions can be extracted by using the approach developed by Faessler and collaborators \cite{khoa1992a,khoa1992b,puri1992}. From a microscopic picture of two
interpenetrating pieces of nuclear matter,  thermal quantities are deduced from the matter density and kinetic energy density. The extraction of the thermal properties is based on a generalized hot Thomas-Fermi formalism (GHTFF), more detailed information can be found in Ref.~\cite{barranco1981,rashdan1987}.

\begin{figure}
\includegraphics[width=8.5cm]{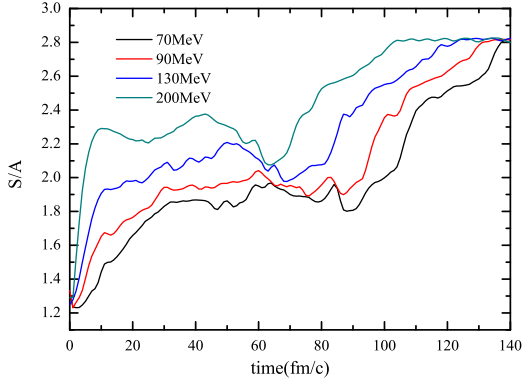}
\vspace{-0.1truein} \caption{\footnotesize    Time evolution of local  entropy per nucleon generated in the central region. Different color line represents different beam energy as  illustrated in the inset.}\label{fig_sa}
\end{figure}

Time evolutions of  temperature and entropy density are depicted in Fig.~\ref{fig_temend}. Along the time scale of the collision, one can see that both values almost evolve isochronously, reach their maxima at about 20 fm/c. After the compression stage the nuclear system begins to expand and some nucleons escape from the central region, and the central region becomes cooled down. The entropy density decreases more quickly than temperature, this is due to  direct effect of the quick escape of the nucleons.
As one can see in Fig.~\ref{fig_sa} the entropy per nucleon $S/A$ in the central region almost evolves isentropically after the compressed stage for some times, the higher the incident energy the larger the $S/A$ is. This can be understood by the transition of the energy of collective motion into the colliding system's thermal motion. And then $S/A$ increases again and reaches a saturate value for all  incident energies. Further more it is interesting that although thermal properties differ from each other during compression and expansion stage for different beam energies, the nuclear matter located in the central region has the same thermal properties at the end of evolution.

\begin{figure}
\includegraphics[width=8.5cm]{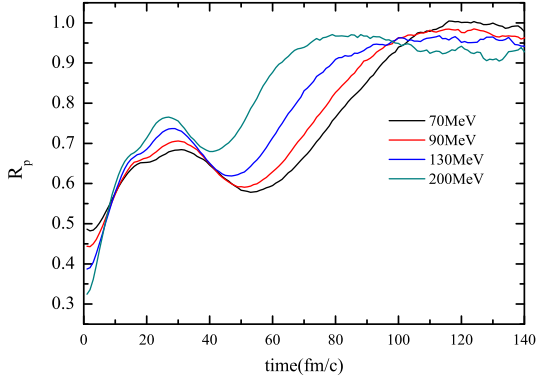}
\vspace{-0.2truein}
\caption{\footnotesize   Time evolution of stopping in the central sphere.  Different color line represents different beam energy as  illustrated in the inset.}
\label{fig_eq}
\end{figure}

Shear viscosity determines the strength of  the energy momentum fluctuation of dissipative fluxes about the equilibrium state, which can be  calculated by using Green-Kubo relation.
The Green-Kubo formula for shear viscosity is defined by \cite{Kubo1966}
\begin{eqnarray}
\label{eq7} \eta = \frac{1}{T} \int{d^{3}r}\int_{0}^{\;\infty}{dt}\langle
{\pi_{ij}}(0,0){\pi_{ij}}({\bf r},t) \rangle,
\end{eqnarray}
where $T$ is the temperature of the system, "0" represent the starting time when the system tends to equilibrium and  $t$ is the
post-equilibration time, $\langle {\pi_{ij}}(0,0){\pi_{ij}}({\bf r},t) \rangle$
is the shear component of the energy momentum tensor. In this work, the post-equilibration stage is defined as the nuclear matter within the given central region has reached  an equilibrium which can be judged by the stopping parameter \cite{Zhang}. The stopping $R_{p}$ is defined as
\begin{equation}
R_{p}=\frac{2\sum R_{t}}{\pi \sum R_{z}},
\end{equation}
where $R_{t}=\sqrt{p_x^{2}+p_y^{2}}$ and $R_{z}=\sqrt{p_z^{2}}$ is the transverse and parallel momentum, resepectively. Time evolution of stopping is displayed in Fig.~\ref{fig_eq}. From the figure,  we observe that at the the initial stage the stopping  is very small and its value decreases with the energy. Later there is a quick increase of the stopping due to the interplay of two body collision, mean field and Pauli blocking, and the collective motion energy turns into thermal energy and the potential energy during the compression stage. It is interesting to see there is a vibration of $R_{p}$, this is caused by the transformation between the potential energy and kinetic energy. The stopping approaches a saturate value very close to 1 after oscillation, which means the nucleonic system in the central volume is very close to equilibrium in later stage of collisions. Different statring time ("0") when $R_p$ tends to 1 has been used in the Eq.~(\ref{eq7}) for calculating viscoisty.

\begin{figure}
\includegraphics[width=8.5cm]{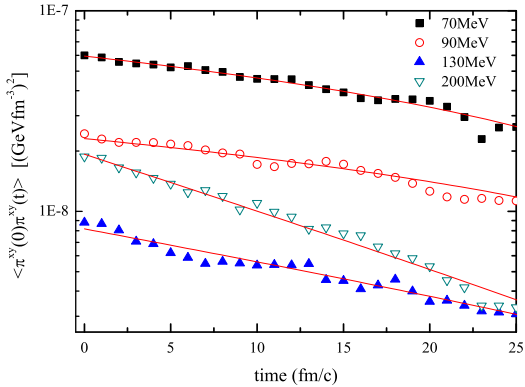}
\vspace{-0.1truein} \caption{\footnotesize  $\langle {\pi_{ij}}(0,0){\pi_{ij}}({\bf r},t) \rangle$ evolves with the post-equilibration time for
the head-on Au+Au collision in the central sphere. Different symbols represent different beam energies and lines are the fits with Eq.~\ref{rex}.
}\label{fig_tcorre}
\end{figure}

The expression for the energy momentum tensor is defined by
$\pi_{ij}=T_{ij}-\frac{1}{3}{\delta}_{ij}T_{i}^{i}$ where the momentum
tensor reads \cite{Muronga2004}
\begin{eqnarray}
\label{eq8}
T_{ij}({\bf r},t)=\int{d^{3}p}\frac{p^{i}p^{j}}{p^{0}}f({\bf r},{\bf p},t),
\end{eqnarray}
with the momenta component $p^i$, $p^j$ and the total energy $p^0$ of each nucleon
provided by the IQMD model, $f({\bf r},{\bf p},t)$ is the phase space density of the particles in the system. In order to compute an integral, we assume that
nucleons are uniformly distributed inside the volume. Meanwhile, the
 volume with the radius $R$=3.5fm is fixed, so the
viscosity becomes
\begin{eqnarray}
\label{eq9} \eta = \frac{V}{T} \langle {\pi_{ij}(0)}^{2} \rangle{\tau}_{\pi},
\end{eqnarray}
where $\tau_{\pi}$ represents relaxation time and can be extracted from the fit:
\begin{equation}
\langle {\pi_{ij}(0)}{\pi_{ij}(t)} \rangle \propto \exp{(-\frac{1}{\tau_{\pi}})} \label{rex}.
\end{equation}

\begin{figure}
\includegraphics[width=8.5cm]{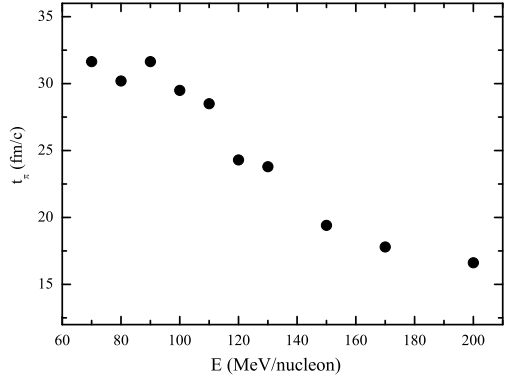}
\vspace{-0.1truein} \caption{\footnotesize    The relaxation time  as a function of incident energy for the head-on Au+Au collision for the central sphere.
}\label{fig_trelax}
\end{figure}

As shown in Fig.~\ref{fig_tcorre}, $\langle {\pi_{ij}}(0,0){\pi_{ij}}({\bf r},t) \rangle$ is plotted versus time for Au + Au collision at different incident energies.
The correlation function is damped exponentially with
time and can be  fitted by the Eq.~(\ref{rex}) to extract the
inverse slope which corresponds to the relaxation time.  Fig.~\ref{fig_trelax}
shows that the relaxation time decreases as
incident energy, indicating that the system can approach to
equilibration faster at higher incident energy, which is consistent with the results from the stopping parameter.

\begin{figure}
\includegraphics[width=8.8cm]{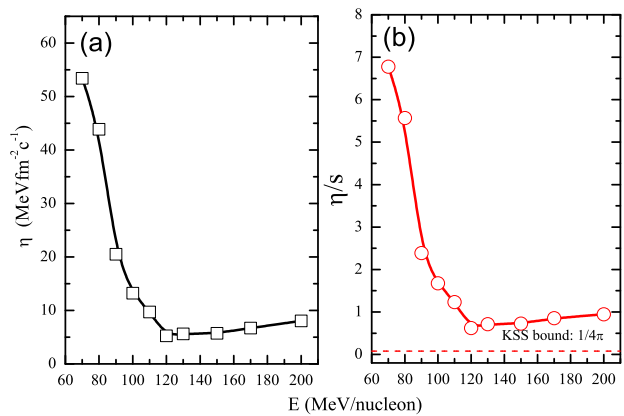}
\vspace{-0.1truein} \caption{\footnotesize Shear viscosity $\eta$ (a) and the ratio of shear viscosity to entropy density (b) as a function of incident energy for the head-on Au + Au collisions.}
\label{fig_eta}
\end{figure}

Finally,  we come by the shear viscosity and its ratio to  entropy density.  Fig.~\ref{fig_eta}(a)  shows shear viscosity $\eta$ as a function of incident energy. It is very interesting to see that the shear viscosity alone also exhibits a minimum near 120 MeV/nucleon. Since the temperatures in post-equilibration stage  are almost the same for the collisions at different energies,  the value of shear viscosity are mostly influenced by the tensor correlation, which reflects the fluctuation of dissipative fluxes.
Fig.~\ref{fig_eta}(b) displays $\eta/s$ as a function incident energy for the head-on Au + Au collisions  from 70  to 200 MeV/nucleon. The $\eta/s$ decreases quickly with the incident energy up to a platform of the minimum value at around 120 MeV/nucleon and afterwards the curve weakly rise up  slowly.  The minimum value is around 0.6, i.e. about 7 times KSS bound ($1/4\pi$),  which is very close to the saturated value in our previous BUU calculation \cite{Li2011}.   As we expected, all $\eta/s$ values from the present QMD model are  lager than the KSS bound.  On the other hand, we noticed that the location of energy with the minimum $N_{IMFs}$ and $\eta/s$ is not exactly the same, it might be due to our Green-Kubo calculation  focuses on  a given central nuclear region, and the IMFs represent the whole colliding system.

\section{\label{sec:level4}IV. Conclusion}

In this Letter, we performed a first calculation on the ratio of shear viscosity to entropy density in the framework of quantum molecular dynamics model with the Green-Kubo method.  For an example,  the head-on Au+Au collisions are simulated  for beam energies from 70 to 200 MeV/nucleon. The generalized hot Thomas-Fermi formalism is used to extract the thermal properties of the given central nuclear system, and the shear viscosity is calculated with the Green-Kubo relation. The results display that a minimum $\eta/s$ region,   about 7 times Kovtun-Son-Starinets (KSS) bound (1/4$\pi$),  exhibits at around  120 MeV/nucleon, which may correspond to a  liquid gas phase transition for a hot nuclear matter. The ratio of shear viscosity to entropy density is always larger than $1/4\pi$, supporting the result of AdS/CFT.
Finally, we would mention that the effect of equation of state and symmetry energy on the $\eta/s$ is also an interesting topic  \cite{tsang2004,Souza2008,Wu2011,Jiang}, which is beyond 
the scope of this Letter,  will be studied  in near future.

\section{ACKNOWLEDGMENTS}
This work is partially supported by the NSFC under contracts No.
11035009, No. 10979074, No. 10875160, No. 10805067 and No. 10975174,  the Knowledge Innovation Project of Chinese
Academy of Sciences under Grant No. KJCX2-EW-N01.

\end{document}